\documentclass[preprint]{aastex}
\shorttitle{Optimizing PTAs}
\shortauthors{Burt et al.}

\begin{document}

\title{
Optimizing Pulsar Timing Arrays to Maximize Gravitational Wave Single Source Detection:  a First Cut}

\author{
Brian J. Burt\footnote{Franklin \& Marshall College, Department of Physics and Astronomy, Lancaster, PA 17604}~\footnote{Currently at Northern Arizona University, Department of Physics and Astronomy, Flagstaff, AZ 86011} ,
Andrea N. Lommen\footnotemark[1] ,
Lee S. Finn\footnote{Pennsylvania State University, Department of Physics, Department of Astronomy \& Astrophysics, University Park, PA 16802} 
}

\medskip

\begin{abstract}

Pulsar Timing Arrays (PTAs) use high accuracy timing of a collection of low timing noise pulsars 
to search for gravitational waves in the microhertz to nanohertz
frequency band. The sensitivity of such a PTA
depends on (a) the direction of the gravitational wave source, (b) the timing accuracy of
the pulsars in the array and (c) how the available 
observing time is allocated among those
pulsars. Here, we present a simple way to calculate the sensitivity of the PTA as a function
of direction of a single GW source, based only on the location and root-mean-square residual
of the pulsars in the array.
We use this calculation to
suggest future strategies for the current North American Nanohertz
Observatory for Gravitational Waves (NANOGrav) PTA in its goal of detecting single GW sources. We also
investigate the affects of an additional pulsar on the array
sensitivity, with the goal of suggesting where PTA pulsar searches
might be best directed. 
We demonstrate that, in the case of single GW sources, if we are interested in maximizing the volume of space
to which PTAs are sensitive, 
there exists a slight advantage to finding a new
pulsar near where the array is already most sensitive.
Further, the study suggests that more observing time
should be dedicated to the already low noise pulsars in order to have
the greatest positive effect on the PTA sensitivity. 
We have made a web-based sensitivity mapping tool available at \url{http://gwastro.psu.edu/ptasm}.

\end{abstract}

\section{Introduction}

Pulsar Timing Arrays (PTAs) \citep{1990FosterBacker}
are a practical means of observing gravitational waves (GWs)
associated with supermassive ($>10^8M_{\odot}$) black holes, or more
generally GWs in the nanohertz to microhertz regime 
arising from any source \citep{jb03,jhv+06, 2010SesanaVeccio, Olmez10}.  
A PTA is a collection
of low timing noise pulsars timed to high accuracy.  
In such a timing array, GWs may signal their presence
through correlated disturbances in measured pulse arrival times. Over
the last decade, assessments of expected GW source strengths have
remained steady \citep{2010SesanaVeccio} while the number of pulsars
that can be timed to high precision has increased, 
and instrumentation, observing technique, and timing precision of
known pulsars have all improved \citep{Hobbs10}. Detection of GWs via pulsar
timing is now within striking distance. Properly made strategic
choices for observing and/or improving the timing precision of current
array pulsars, and searching for new pulsars, can hasten the arrival of the day when
the first detection is made and the field of GW astronomy is properly
inaugurated. 

Strategic optimization depends upon your strategy.  
Do you want to detect single sources, or a stochastic background?  Do you
want to maximize the number of sources detected, or maximize the signal-to-noise
ratio (SNR) of the sources that you do detect?
Here we investigate the 
sensitivity of a PTA to a single source of GWs and
investigate optimization strategies that maximize the number of sources detected.
We describe the results of a preliminary
investigation into strategies for deploying limited observing time
among current array pulsars, and for searching for new pulsars to
augment the current array.  Our results apply to the search for
single sources of GWs\citep{Finn10, Yardley10}, not to stochastic background searches
\citep{Jenet06, Anholm08, vanHaasteren09}.

After an introduction of the PTA single-source
sensitivity calculation and NANOGrav characteristics in \S 2, we
discuss in \S 3 the consequences of adding pulsars to the
array. Further, discussed in \S 4, is the preliminary study of
pulsar timing allocation time. Finally, in \S 5, we discuss the
implications that rise from this study and suggest future research.

\section{Calculating PTA Sensitivity}

Anticipated GW sources in the nHz-$\mu$Hz band include nearby
supermassive black hole binary and triplet systems \citep{jb03, 2010SesanaVeccio, Amaro-Seoane10}. More
speculative sources include a primordial stochastic background and
bursts from cosmic string cusps and kinks \citep{Damour01, damour2005, Siemens06, Siemens07, Olmez10}. All these sources
are expected to be isotropically distributed on the
sky. Correspondingly we choose to estimate the sensitivity of PTAs as
the spatial volume within which a fiducial source would give an
SNR greater than a fixed threshold. Such a measure
can be made independent of the threshold (but not the fiducial source)
by referring the calculated volume to a reference volume calculated
for a reference array. In this way we define the relative overall sensitivity
$\nu_{overall}$ of a PTA as:

\begin{equation}
\nu_{overall}=\frac{\sum_{i=1}^n d(\hat{k}_i)^3}{\sum_{i=1}^n d(\hat{k}_i)_{R}^3},
\label{eq:distance}
\end {equation}

\noindent where $d$ is the distance out to which the PTA is sensitive to a source 
propagating in direction $\hat{k}_i$ relative to the reference PTA denoted with a 
subscript $R$. The limit $n$ is the number of directions along which sensitivity is 
being measured and can be chosen by the user according to the desired resolution
of the resulting sensitivity map.  In the maps we constructed in this paper we used
HEALPix\footnote{\url{http://healpix.jpl.nasa.gov}} to construct pixels representing equal surface
area, and $n=3072$.
% n = 12*nSide^2 = 12*16^2 = 3072
% nSide=16 must be used in the final figures.  if nSide=8 is sued then this number is 768.

The distance to a fiducial GW source in a fixed direction is inversely proportional to its amplitude SNR $\rho$ as observed in a PTA. 
Correspondingly, we may use the amplitude SNR 
of a fiducial source (assuming we have measured the optimal SNR) as a surrogate for the distance 
of the source
\begin{eqnarray}\label{eq:r0}
d(\hat{k},\rho) = d(\hat{k},\rho_0)\frac{\rho_0}{\rho},
\end{eqnarray}
where $d(\hat{k},\rho)$ is the distance to the source when its anticipated SNR is $\rho$, its propagation direction is $\hat{k}$, and $\rho_0$ is the amplitude SNR of the fiducial source at distance $d(\hat{k},\rho_0)$. 

Timing noise for typical PTA pulsars is typically white on timescales less than 5--10 years and red on longer timescales 
\citep{jhv+06, Hobbs06}.
The white contribution to the timing noise is characterized by its root mean square (RMS) residual $\sigma_j$. 
For our approximate analysis here we ignore the red timing noise contribution and assume, for each PTA pulsar, white timing noise characterized by $\sigma_j$ for pulsar $j$, and that a single
$\sigma_j$ characterizes the entire observation. 

With this timing noise approximation the contribution from pulsar $j$ to the PTA power SNR in direction $\hat{k}$ is
\begin{eqnarray}\label{eq:rho2j}
\rho^2_j(\hat{k},t) &= \frac{\tau^2_{GW}(\hat{k},t)_j}{\sigma_j^2}, 
\end{eqnarray}
where $\tau_{GW}(\hat{k},t)_j$ is the anticipated GW contribution to the residuals in pulsar $j$ from a GW source propagating in 
direction $\hat{k}$ at time $t$.
The anticipated power SNR $\rho^2$ of the whole array to sources propagating in direction $\hat{k}$ is the sum over the pulsars
\begin{eqnarray}\label{eq:rho2}
\rho^2(\hat{k},t) &= \Sigma_{j=1}^{n_p}\frac{\tau^2_{GW}(\hat{k},t)_j}{\sigma_j^2}, 
\end{eqnarray}
where $n_p$ is the number of pulsars.
Armed with this equation, we set off to find $\tau_{GW}(\hat{k},t)_j$.

The timing residuals $\tau_{GW}$ associated with TT-gauge GW metric perturbation $h_{+}\mathbf{e}_{+}+h_{\times}\mathbf{e}_{\times}$ may be written as
\begin{eqnarray}
\label{eq:tauj}
\tau_{GW}(\hat{k},t)_j &= 
F^{+}(\hat{k},\hat{n}_j) g_{+}(t,L_j,\hat{k}\cdot\hat{n}_j) +
F^{\times}(\hat{k},\hat{n}_j) g_{\times}(t,L_j,\hat{k}\cdot\hat{n}_j),
\end{eqnarray}
where $\hat{n}_j$ is a unit vector
pointing to pulsar $j$, $L_j$ is the distance to that pulsar, and $\hat{k}$ is the direction of propagation of the GW.
$F^{+/\times}$ are given by
\begin{eqnarray}
\label{eq:f+}
F^{+}(\hat{k},\hat{n}_j) &= -\frac{1}{2} \hat{n}_j^T \mathbf{e}^+(\hat{k}) \hat{n}_j,\\
\label{eq:fx}
F^{\times}(\hat{k},\hat{n}_j) &= -\frac{1}{2} \hat{n}_j^T \mathbf{e}^{\times}(\hat{k}) \hat{n}_j,
\end{eqnarray}
where $\hat{n}^T$ denotes the transpose of $\hat{n}$,
and $\mathbf{e}^{(+)}$ and $\mathbf{e}^{(\times)}$ are the two independent gravitational wave polarization basis tensors,
\begin{eqnarray}
e_{(+)}^{lm}e^{(+)}_{lm} = e_{(\times)}^{lm}e^{(\times)}_{lm} = 2\\
e_{(+)}^{lm}\hat{k}_{m} = e_{(\times)}^{lm}\hat{k}_m = e_{(+)}^{lm}e^{(\times)}_{lm} = 0.
\end{eqnarray}
Functions $g_{+}$ and $g_{\times}$ are integrals of $h_{+}$ and $h_{\times}$ as follows \citep{Finn10}: 
\begin{eqnarray}
g_{(+/\times)}(t, L_j, \hat{k}_{j}\cdot \hat{n}^{j})=\int_0^{L_j} h_{+/\times} \left(t-(1+\hat{k}\cdot\hat{n}_j)(L_j-\lambda)\right)d\lambda.
\end{eqnarray}
Note that we are using geometrized units where $c=G=1$.
We have essentially broken up 
$ \tau_{GW}(\hat{k})_j $ 
into terms that depend on geometry ($F$'s) and terms that depend on
time ($g$'s).  Following \citep{Finn10} we assume that a function $f$ exists for which
\begin{equation}
df_{+/\times}(u)/du = h_{+/\times}(u).
\end{equation}
We can then do the integral as follows:
\begin{eqnarray}
\label{earth_and_pulsar_terms}
g_{(+/\times)}(t, L_j, \hat{k}_{j}\cdot \hat{n}^{j})= \frac{f_{+/\times}(t)}{1 + \hat{k}\cdot\hat{n}_j} -  \frac{f_{+/\times}(t - (1 + \hat{k}\cdot \hat{n}_j)L_j)}{1 + \hat{k}\cdot\hat{n}_j}.
\end{eqnarray}
\noindent The first term is the so-called `earth term', and the second the `pulsar term'\citep{2004JenetLommen}.
The pulsar term is delayed from the earth term by $(1 + \hat{k}\cdot \hat{n}_j)L_j$ which amounts to hundreds to thousands of years in most cases.
For the moment,
assume that we are dealing with burst sources whose length is shorter than our observation time (years) 
for which we only observe the earth term, and that we can thereby ignore the pulsar term.
Later we will show that the result we derive
here holds for continuous sources, when the pulsar term must be included, as well.

$ \tau_{GW}(\hat{k},t,t)_j $ 
depends on time, so what we desire is the time averaged SNR, $\overline{\rho^2}$, i.e. the time average of equation 
\ref{eq:rho2}.  We square equation \ref{eq:tauj} and average over time and all possible polarizations $h_{+}$ and $h_{\times}$. 
The cross-term 
$F^{+}(\hat{k},\hat{n}_j) F^{\times}(\hat{k},\hat{n}_j)$ vanishes when we average over all polarizations.
Also by averaging over all polarizations we find
$\overline{g_{+}^2}=\overline{g_\times^2}$. 
Finally, exploiting time translation symmetry we can write
\begin{eqnarray}\label{eq:rho2barj}
\overline{\rho^2_j} &= A^2\left(\frac{F^{(+)2}(\hat{k},\hat{n}_j)+F^{(\times)2}(\hat{k},\hat{n}_j)}{\sigma_j^2(1 + \hat{k}\cdot \hat{n}_j)^2}\right),
\end{eqnarray}
where $A^2$ is a constant independent of the pulsar line-of-sight $\hat{n}_j$ or the propagation direction of the GWs $\hat{k}$. 

Combining equations \ref{eq:r0}, \ref{eq:rho2}, \ref{eq:f+}, \ref{eq:fx} and \ref{eq:rho2barj}  
we have our principal result
\begin{eqnarray}
\label{principal_result}
\overline{\rho^2(\hat{k})} &= A^2\sum_{j=1}^{n_p}\frac{F^{(+)2}(\hat{k},\hat{n}_j)+F^{(\times)2}(\hat{k},\hat{n}_j)}{\sigma_j^2(1 + \hat{k}\cdot \hat{n}_j)^2},\\ 
                           &= \frac{A^2}{4}\sum_{j=1}^{n_p}\frac{\left(1 - (\hat{k}\cdot \hat{n}_j)^2\right)^2 }{\sigma_j^2 (1 + \hat{k}\cdot\hat{n}_j)^2}, \\
                           &= A^{\prime 2}\sum_{j=1}^{n_p}\left(\frac{1 - \hat{k}\cdot \hat{n}_j}{\sigma_j}\right)^2.
\end{eqnarray}
The steps between equations 14 and 15 can be done for any $\hat{k}$, but the result in equation
15 can be seen more readily by assuming the GW is traveling in the $\hat{z}$ direction which
gives
\[ \mathbf{e}^+(\hat{k}) = \left| \begin{array}{ccc}
1 & 0 & 0 \\
0 & -1 & 0 \\
0 & 0 & 0 \end{array} \right|,\] 
\[ \mathbf{e}^\times(\hat{k}) = \left| \begin{array}{ccc}
0 & 1 & 0 \\
1 & 0 & 0 \\
0 & 0 & 0 \end{array} \right|,\]
and assuming an arbitrary pulsar direction $\hat{n}_j = [n_x, n_y, n_z]$.
The numerator in the sum in equation 14 becomes (after some algebra and trigonometry)  
$ (1 - n_z^2)^2/4$.  
For an arbitrary GW direction, $\hat{k}$, the numerator generalizes to $(1-(\hat{k}\cdot\hat{n}_j)^2)^2/4$.
In other words, the quantity that matters is the projection of the pulsar direction vector
onto the direction of propagation of the GW.

Equation 16 will allow us to compare various PTAs independently of the details of the input source ($g_+$ and $g_\times$ as shown in equation
\ref{eq:tauj})
and then putting this into equation \ref{eq:distance}:
\begin{eqnarray}
\label{eq:nu}
\nu_{overall} &= \frac{\Sigma_{i=1}^n \overline{\rho^2(\hat{k_i})}^{3/2}}{\Sigma_{i=1}^n \overline{\rho_R^2(\hat{k_i})}^{3/2}}, 
\end{eqnarray}
where $\overline{\rho_R^2}$ refers to the reference PTA, and, as in equation \ref{eq:distance}, the $n$ is the resolution of the calculation, i.e.
the number of pixels in the map of $\overline{\rho^2(\hat{k_i})}$.  $\nu_{overall}$ is what we are calling the `volume sensitivity' and represents
the ratio of the volumes to which two different PTAs are sensitive.  

A related quantity which we will utilize later, ${\nu}(\hat{k})$, represents
the comparison of the sensitivity of two arrays as a function of the GW propagation direction $\hat{k}$, 
\begin{eqnarray}
\label{eq:nuk}
\nu(\hat{k}) &= \frac{ \overline{\rho^2(\hat{k})}^{3/2}}{\overline{\rho_R^2(\hat{k})}^{3/2}}. 
\end{eqnarray}
Note that rather than plotting $\nu(\hat{k})$ as a function of GW propagation direction $\hat{k}$ we will plot $\tilde{\nu}(\hat{s})$ where
$\hat{s}$ is the direction of the GW source in the sky, $\hat{s} = -\hat{k}$, and $\tilde{\nu}(\hat{s}) = \nu(-\hat{k})$.

The quantity 
\begin{eqnarray}
\label{eq:P2}
P^2(\hat{k}) &= \frac{\overline{\rho^2(\hat{k})}}{\max_{\hat{k}}\overline{\rho^2(\hat{k})}}
\end{eqnarray}
lends itself to interpretation as the PTA \emph{antenna pattern:} i.e., it is proportional to the signal power absorbed by the detector 
from a source propagating in the direction $\hat{k}$, measured relative to the source direction for which the greatest power is absorbed. 
As with $\nu(\hat{k})$, $\tilde{P^2}(\hat{s}) = P^2({-\hat{k}})$.
Figure 1
shows ``source-averaged'' antenna pattern $\tilde{P}^2(\hat{s})$ for the NANOGrav PTA, whose member pulsars and their RMS timing 
residuals at the time of writing 
are provided in Table 1. 
We have also made available a web-based tool for computing sensitivity maps for an arbitrary array of pulsars at this
URL\footnote{\url{http://gwastro.psu.edu/ptasm/}}.

\begin{figure}[!ht]
\centering
%\epsscale{.75}
\plotone{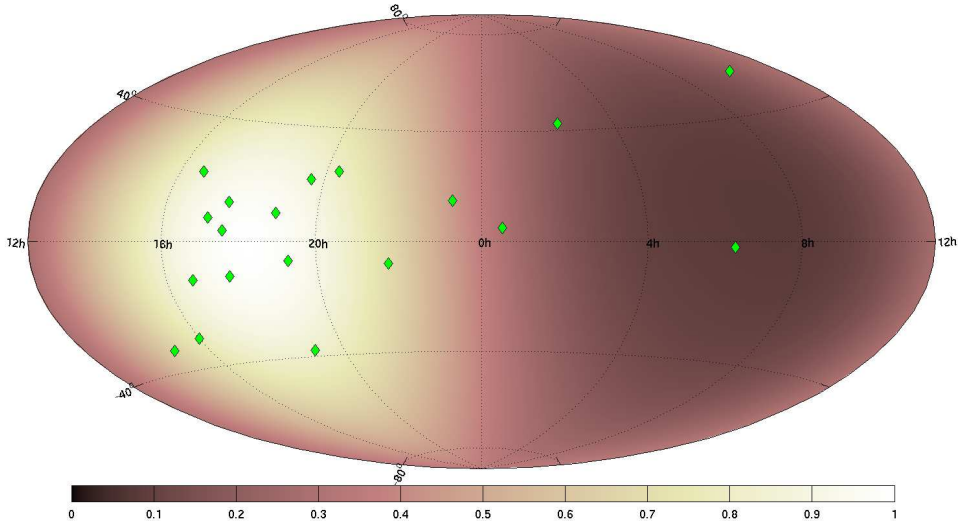}
\caption{Antenna pattern $\tilde{P}^2(\hat{s})$ (defined in the text following equation \ref{eq:P2}) of the NANOGrav PTA
is shown in equatorial coordinates. 
In other words,
this shows the sensitivity of the NANOGrav PTA  to single GW sources as a function of location of the source.
The diamonds show the location of the pulsars in the array.}
\label{fig:NANO}
\end{figure}

As evident from Figure~\ref{fig:NANO}
the sensitivity of the NANOGrav PTA is heavily
biased to the region between 12 and 24 hours right ascension, yet
fairly symmetric about the equatorial plane. Not surprisingly this region
represents both
the highest concentration of pulsars and also the location of
the lowest noise pulsars.

Rather than average over all possible GWs propagating in direction $\hat{k}$ we can restrict attention to any particularly 
interesting class of sources. Consider, for example, the radiation from a circular binary consisting of two supermassive black holes. For 
systems like these the radiation is approximately periodic over any reasonable observational timescale \citep{Jenet06, Sesana10} 
and \begin{equation}
 \tau_{GW}(\hat{k})_j 
\propto \cos\omega t - \cos\omega\left[t - (1-\hat{k}\cdot\hat{n}_j)L_j\right],
\end{equation} 
where 
$\omega$ is the GW angular frequency. Even in the best of circumstances pulsar distances are known to no better than 10\% \citep{Cordes02}, 
in which case the phase $\omega L_j$ is uncertain by many times $2\pi$ for $\omega$ of interest. If we average 
$ \tau_{GW}(\hat{k})_j $ 
over the 
typical uncertainty in distance we find the contribution owing to the term involving $L_j$, the pulsar term, vanishes and we are left with $\cos^2\omega t$ 
which, averaged over time, tends to 0.5, a constant. \emph{As long as the uncertainty in pulsar distance is greater than the 
light travel time over the duration of the observation these same considerations will hold for any source that a PTA can detect:} i.e., 
averaged over the uncertainty in pulsar distances the pulsar term contribution to $\tau^2$ will vanish and 
we can again ignore the second term on the right-hand side of equation \ref{earth_and_pulsar_terms} as we did to obtain the result shown in equations 16.
Note that in the case of $\hat{k}\cdot\hat{n} = 1$ the earth and the pulsar terms exactly cancel and the
sensitivity $\overline{\rho^2(\hat{k})}=0$ as is shown in equation \ref{principal_result}.

\section{Addition of Pulsars}
\label{add_a_pulsar}
Adding a new pulsar to an existing array increases the array's overall sensitivity.  
Equation \ref{eq:nu} shows how the array's sensitivity increases as a function of a new pulsar's sky location and timing residual noise RMS.  
While we do not have the freedom to choose where we will find the next good millisecond pulsar, 
we do have the freedom to choose where we will look.  With this in mind we consider how the sensitivity of the 
NANOGrav PTA would be increased by the addition of a single pulsar whose timing residual noise of 200 ns RMS 
is equal to the current array's median.

\begin{figure}[!ht]
\centering
\epsscale{.75}
\plotone{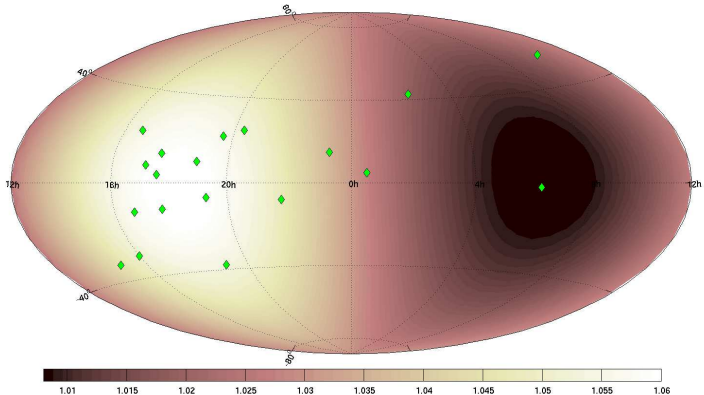}
\caption{Shown, the relative sensitivity improvement $\nu_{overall}$ (equation \ref{eq:nu}) of the NANOGrav PTA caused by the addition of a pulsar, as a function of the location of that added pulsar, in equatorial coordinates. Every pixel represents the relative sensitivity improvement caused by an additional pulsar at that location in the sky.}
\label{fig:AddWhereComp}
\end{figure}

Figure
~\ref{fig:AddWhereComp}
shows the sensitivity $\nu_{overall}$ as calculated using equation \ref{eq:nu} as a function of the location of an addition to the NANOGrav PTA of a single pulsar
with 200 ns timing noise RMS.
As is confirmed by the figure, an
additional pulsar will improve the PTA sensitivity regardless of its
location on the sky, but the improvement
represents less than a 6\% increase in sensitivity volume (the volume of space
from which we can detect sources) in all cases. 
However, some pulsar locations will improve it
more than others. An additional pulsar
in the region in which the PTA is already most sensitive improves the
sensitivity volume the greatest. This seems appropriate considering
that the volume of sensitivity goes as $d^3$. So,
if a distance that is already large is doubled, the volume increase
will be a larger factor than if a small distance were doubled. 
However, the volume sensitivity 
varied by only 6\% as we moved the additional pulsar all over the sky, so it may be wisest to 
search where one is most likely to find pulsars,
such as in the galactic plane.  
One aspect which has not been addressed in this manuscript is the coherence of
the GW signal between pulsars, i.e. the fact that pulsars in similar directions
in the sky will show higher correlation between their GW signals than those
in different parts of the sky (see eq. \ref{earth_and_pulsar_terms}).  In addition, in order to
confirm that the detected signal is a GW and not, for example, an error in earth's ephemerides,
or a terrestrial clock, we will need pulsars in different parts of
the sky to confirm that the spatial signature of the detected signal is quadrupolar in nature and not
dipolar (ephemerides error) or monopolar (clock error).
Further study
which includes these considerations
is necessary to determine the optimum
strategy.  

\section{Optimization With Time Constraints}

The NANOGrav PTA at the time of writing involves 19 pulsars whose timing residuals range from 54 ns to 2.2 $\mu$s.  
Monitoring each pulsar requires some fraction of the available observing time, which is a 
valuable resource.  How should the available observing time be distributed among the different 
pulsars to optimize the NANOGrav PTA's sensitivity?  To explore this question we divide the NANOGrav 
PTA pulsars into ``low-noise" (timing residual noise less than 200 ns) and``high-noise" (timing 
residual noise greater than 200 ns) groups, and evaluate the array sensitivity when we increase the 
fraction of observing time spent on low-noise pulsars at the expense of high-noise pulsars, and vice versa.

\begin{figure}[!ht]
\centering
%\plotone{ImproveWhat.eps}
%
% To make these you say:
% PTASMdist('worst_better.txt', 'nanograv.txt','true', 'eq', '8', 'worst_better.png')
% For "worst" make sure the middlepsr MarkerFace and Color are w  (white)
% For "best" make sure the middlepsr MarkerFace and Color are k (black)
% PTASMdist('best_better.txt', 'nanograv.txt','true', 'eq', '8', 'best_better.png')
% "save" them to eps.
% Change bounding box in box to 141 247 469 504  (otherwise they're too small)
% 247 used to be 287 but then I added the vskip in combination with the 247 to
% make it not delete text anymore
%
\plottwo{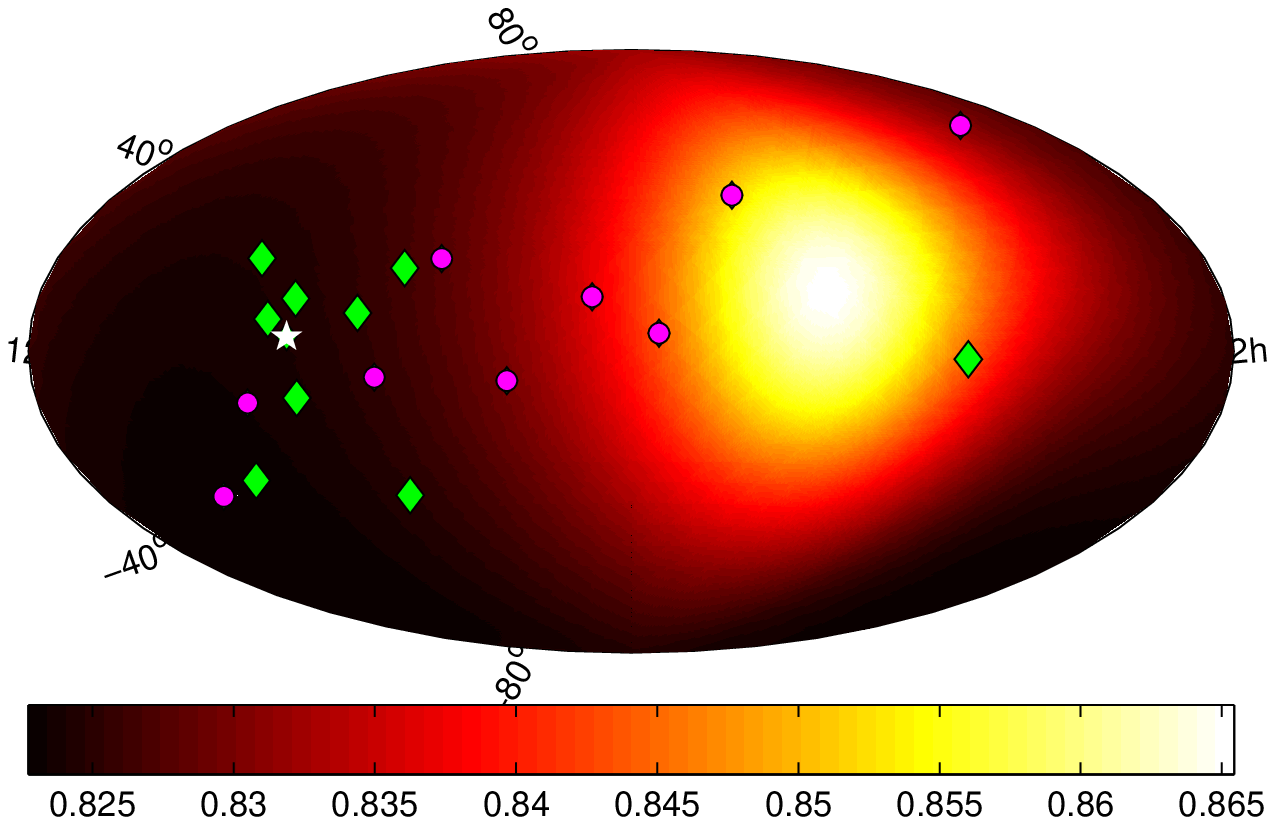}{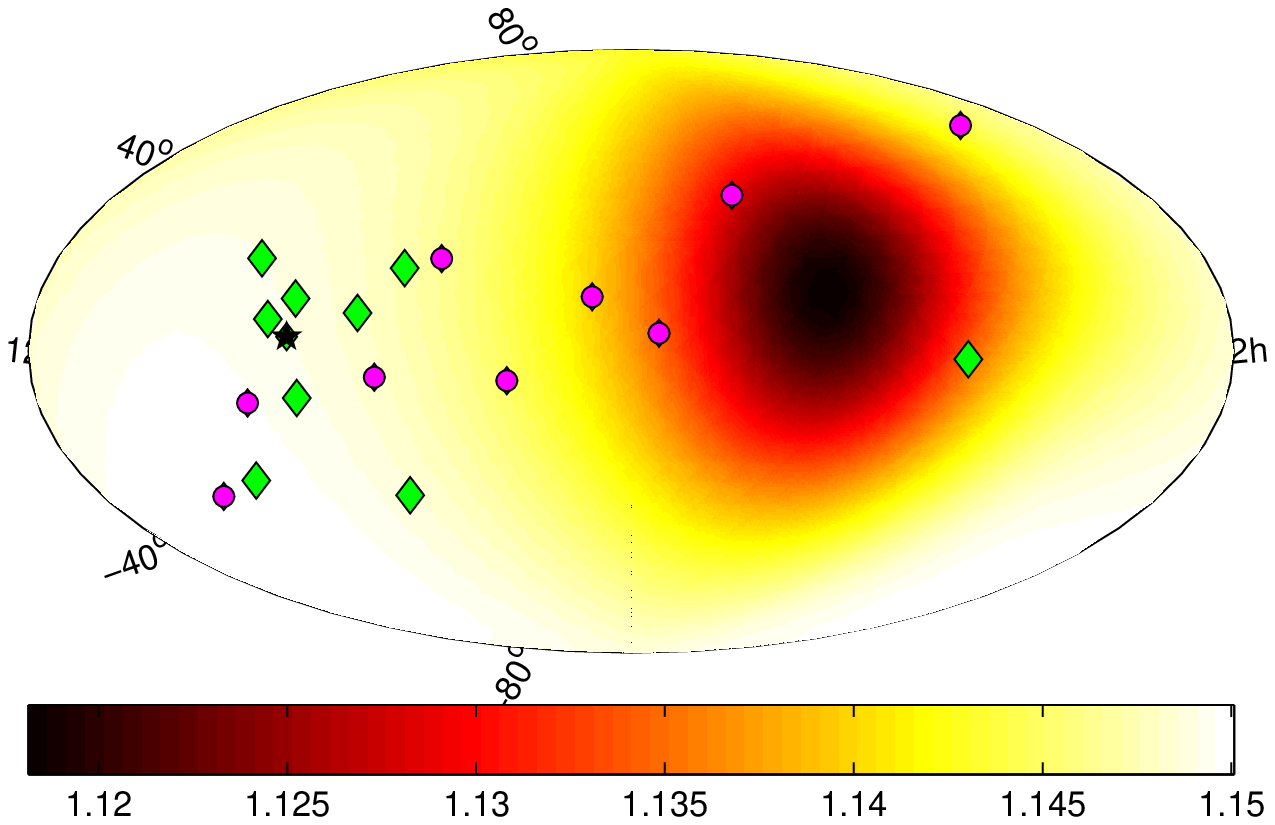}
\caption{Sensitivity improvement compared to the current NANOGrav pulsars and current NANOGrav pulsar RMS timing values
as a function of GW propagation direction $\hat{s}$. 
In both panels we plot 
$\tilde{\nu}(\hat{s})$ as defined by equation \ref{eq:nuk} and the text immediately thereafter.
On the left we improve the RMS of the high-noise pulsars and obtain a volume sensitivity, $\nu_{overall}=0.6$.
On the right we improve the RMS of the low-noise pulsars and obtain a volume sensitivity, $\nu_{overall}=1.5$.
Diamonds are the low-noise pulsars, circles are the high-noise pulsars.  The median pulsar whose RMS we
did not alter is a pentagram.  All coordinates are equatorial.
Please note the scales of the two diagrams are different to allow the reader to see the structure of the plots.}
\label{fig:ImproveWhat}
\end{figure}

For a typical timing array pulsar the RMS timing residual noise, $\sigma$, is inversely proportional 
to the square-root of the time spent observing the pulsar, i.e.,
$\sigma\propto t^{-1/2}$ \citep{2009Verbiest}.
The current NANOGrav timing program spends an approximately equal amount of time on each pulsar in the array.  
We consider two alternatives: spending twice as much observing time on pulsars in the high-noise group 
as on pulsars in the low-noise group, and the opposite case of twice as much observing
time on pulsars in the low-noise group as on pulsars in the high-noise group.  
This translates to the observing time of one group
being multiplied by $\frac{4}{3}$ while the observing time of the
other group is multiplied by $\frac{2}{3}$. This, as RMS and $t$ are
related above, results in an decrease in RMS of the first group by
$\sqrt{3/4}$ and an increase in RMS of the second group by
$\sqrt{3/2}$.
(In
both cases the time spent observing the single pulsar with timing residual noise of exactly 200 ns is left fixed.)  
Figure ~\ref{fig:ImproveWhat}
shows the change in volume sensitivity enacted by these adjustments as a function of GW propagation direction, $\hat{s}$.
%$\tilde{\nu}(\hat{s})_{adjusted}/\tilde{\nu}(\hat{s})_{original}$ 
What we plot is
$\tilde{\nu}(\hat{s})$
where the reference array in the denominator (see equation \ref{eq:nuk}) consists of the NANOGrav pulsars
shown in Table \ref{table:PulsarList}.  The array used in the numerator is one in which
the adjustments described above have been assumed.

\begin{table}[!ht]
  \centering
   %\begin{tabular}{ l c c c}
   \begin{tabular}{ l c c }
    \multicolumn{3}{c}{NANOGrav Pulsars\footnotemark[5]}\\
    \hline
    \hline
      & Pulsar & RMS($\mu$s) \\% & Distance(kpc) \\
    \hline
    1 & J0030+0451	& 0.300    \\%& 0.24\\
    2 & J0218+4232      & 0.830    \\%& 5.85\\
    3 & J0613-0200	& 0.110    \\%& 0.48\\
    4 & J1012+5307      & 0.540    \\%& 0.52\\
    5 & J1455-3330	& 0.960    \\%& 0.74\\
    6 & J1600-3053	& 0.190    \\%& 2.67\\
    7 & J1640+2224      & 0.110    \\%& 1.19\\
    8 & J1643-1224	& 1.100    \\%& 4.86\\
    9 & J1713+0747	& 0.055    \\%& 1.05\\
    10 & J1738+0333	& 0.200    \\%& 1.97\\
    11 & J1741+1300	& 0.140    \\%& 1.00\footnotemark[5]\\
    12 & J1744-1134	& 0.130    \\%& 0.48\\
    13 & J1857+0943     & 0.066    \\%& 0.91\\
    14 & J1909-3744	& 0.054    \\%& 1.14\\
    15 & J1918-0642	& 0.960    \\%& 1.40\\
    16 & J1939+2134	& 0.080    \\%& 8.33\\
    17 & J2019+2425	& 0.910    \\%& 0.91\\
    18 & J2145-0750	& 0.750    \\%& 0.50\\
    19 & J2317+1439	& 0.369    \\%& 1.89\\
    \hline
    \multicolumn{3}{c}{PPTA Pulsars not part of NANOGrav\footnotemark[6]}\\
    \hline
    20 & J0437-4715      & 0.10  \\
    21 & J0711-6830      & 1.00  \\
    22 & J1022+1001      & 0.50  \\
    23 & J1024-0719      & 1.00  \\
    24 & J1045-4509      & 1.00  \\
    25 & J1603-7202      & 0.50  \\
    26 & J1730-2304      & 1.00  \\
    27 & J1732-5049      & 1.00  \\
    28 & J1824-2452      & 1.00  \\
    29 & J2124-3358      & 1.00  \\
    30 & J2129-5721      & 1.00  \\
    31 & J2145-0750      & 0.30  \\
    \hline
  \end{tabular}
\caption{The NANOGrav pulsars with the current RMS timing values at the time of writing.
$^5$This is a particular characterization of these 
NANOGrav pulsars based on communications at the time of this writing with the 
North American 
Nanohertz Observatory for Gravitational Waves.  It is not a definitive characterization.  We are not presenting the data 
associated with these pulsars but rather using them as an example of a realistic PTA.
$^6$RMS values for EPTA pulsars from \cite{Hobbs09}.
}
\label{table:PulsarList}
\end{table}

As is noticeable from Figure ~\ref{fig:ImproveWhat} (with two
scales required for plot structure), bettering the already good
pulsars improves the overall array sensitivity volume by a factor of
1.5 ($\nu_{overall} = 1.5$, see equation \ref{eq:nu} for definition of $\nu_{overall}$), while bettering
the bad pulsars has quite the opposite effect actually worsening the
current sensitivity volume ($\nu_{overall} = 0.6$). 
The key to understanding this result is in noting that SNR $\rho$, residual response $\tau$ and
RMS $\sigma$ are related to each other approximately as follows
\begin{equation}
\rho \propto \frac{\tau}{\sigma},
\end{equation}
and $\tau = \tau_0(d_0/d)$ where $\tau_0$ is the amplitude of the residual at earth 
when the source is at distance $d_0$.  Therefore
\begin{equation}
\rho \propto \frac{\tau_0}{\sigma d},
\end{equation}
or for a fixed SNR $\rho$ required for detection, the distance out to which we could
detect a source, $d$, is inversely proportional to the RMS $\sigma$
\begin{equation}
d \propto 1/\sigma.
\end{equation}
The argument then is similar to that which we made in \S 3, that volume sensitivity goes as $d^3$ 
so halving an already small RMS $\sigma$
increases the volume sensitivity by a much larger factor than halving a larger RMS $\sigma$.
\vskip 0.5 in

We can use the same formalism to determine the value of adding more pulsars to the array and can
use it to ascertain the value of the International Pulsar Timing Array (IPTA) a collaboration
formed of NANOGrav, the PPTA and the the European Pulsar Timing Array (EPTA).
In table \ref{table:PulsarList} below
the NANOGrav pulsars we have shown the Parkes Pulsar Timing Array pulsars, not including those overlapping with NANOGrav,  
as published in the status paper by \citet{Hobbs09}.  
If solely PSR J0437-4715 is added to the NANOGrav array, the volume sensitivity increases by 7\%.  If all the PPTA
pulsars listed are added to the NANOGrav array, the volume sensitivity increases by 10\%.  If instead we imagine
an improved situation in which the PPTA pulsars listed are added with 200ns RMS, with the exception of PSR J0437-4715 which we
add at its actual RMS of 100 ns, the volume sensitivity is
increased by 35\%, ie 35\% more volume of space is sampled for the same GW source type.
At the time of writing the European Pulsar Timing Array  
(EPTA) pulsars for which RMSs are available\citep{Ferdman10} are already included in this list.  
Their typical RMSs are slightly higher than the values shown here, but there are several reasons
to expect that within a year the EPTA RMSs will be markedly reduced and that the pulsars will contribute significantly to this list.
First, the Large European Array of Pulsars (LEAP) project which expects first light in late 2010 
will create a ``tied-array" mode for the 5 European 100-m class dishes into a single instrument,
rivaling the sensitivity of Arecibo, but with larger sky coverage.  Furthermore, the data
reported on by \citet{Ferdman10} represent new instrumentation at all 5 telescopes, so one can expect significant improvement with
characterization and optimization of those instruments.

\section{Discussion and Conclusion}

Given the goal to directly detect GWs, the optimization of pulsar
timing arrays should be well understood. 
Figure~\ref{fig:AddWhereComp} 
suggests that for the sake of GW detection
we may not need to ``fill in" regions of the sky currently
devoid of PTA pulsars, but that rather a clustering of good
pulsars in one region yields the gratest number of detectable GW sources.
The figure shows the sensitivity (volumetric) gain produced by adding a new pulsar 
to the array as a function
of the location of the added pulsar.   
The current pulsars in the array are clustered around 18h right ascension and 0 declination
(as shown by green diamonds on the figure) and in fact the greatest improvement in
the volume to which the array is sensitive is produced by adding a new pulsar near
that already existing cluster of pulsars, although the range of improvement is
modest in all cases (from improving it not at all to improving it by 6\%).
Here we suggest
that if the goal is to maximize the volume of space to which we are
sensitive to GW sources the 
location with highest concentration of pulsars is slightly favored over
other locations, but does not make a significant difference.  So perhaps
efforts to ``fill in the gaps'' in the spatial arrangements of PTAs are unfounded.
However, any new pulsar,
regardless of its location, is beneficial to the scientific
community. Continuing research is needed on this particular area to
optimize pulsar searches for PTA goals.

Our initial findings indicate that the sensitivity to burst
and continuous GWs can be significantly improved by observing longer
the pulsars for which we already have good timing values. In other
words, it is found that the intuitive thing to do, observing longer
the pulsars that we do not have good timing values for, significantly
decreases our sensitivity to GWs by almost 50\% (Figure
~\ref{fig:ImproveWhat}). This suggestion taken to an extreme yields
the ridiculous result that it is best to spend all observing time on
a signal pulsar.  The sensitivity plot in this case would be the 
familiar beam pattern of a single pulsar, but if somehow GWs could
be detected with a single pulsar, spending all our time on this
one best pulsar, assuming the RMS reduces as the square-root of
observing time, would in fact maximize the volume
of space to which we are sensitive.   
We discussed in \S \ref{add_a_pulsar} that this strategy neglects 
issues with confirming the detection as distinct from clock, ephemerides, and other errors, so more 
work must be done to optimize observing strategies, but this work clearly 
indicates that convention may not be the best means to a discovery.
In particular, as the PTA is being optimized, what should the
observing strategy be? Is the answer different depending on whether
our aim is to detect single sources or a stochastic
background. Presented here is a crucial first step toward answering
these questions.

\acknowledgments 
This research was carried out at both Franklin and
Marshall College and Pennsylvania State University. We wish to express
gratitude for the use of facilities at each. We also wish to thank
Matt Kinsey for his MATLAB interface to HEALPix, and Hannah Williams
and Diego Menendez for their assistance in developing a web
interface to the sensitivity calculations for an arbitrary PTA.
LSF and ANL gratefully acknowledge the support of National Science Foundation grants
PHY 06-53462 and AST CAREER 07-48580 respectively.

%\bibliography{references}

\end{document}